\documentclass[prlb,aps,amssym,nofootinbib,floatfix,twocolumn,notitlepage]{revtex4-1} 

 \usepackage{xcolor}
\usepackage{amsmath,amssymb,bbold,bm}
\usepackage{graphicx}
\usepackage{caption}
\usepackage{subcaption}
\usepackage{cancel}
\usepackage{comment}
\usepackage[normalem]{ulem}

\newcommand{\be}{\begin{equation}}
\newcommand{\ben}{\begin{equation*}}
\newcommand{\ee}{\end{equation}}
\newcommand{\een}{\end{equation*}}
\newcommand{\bs}{\begin{split}}
\newcommand{\es}{\end{split}}
\newcommand{\bmx}{\begin{array}}
\newcommand{\emx}{\end{array}}
\newcommand{\bea}{\begin{eqnarray}}
\newcommand{\bean}{\begin{eqnarray*}}
\newcommand{\eea}{\end{eqnarray}}
\newcommand{\eean}{\end{eqnarray*}}
\newcommand{\dg}{^{\dagger}}
\newcommand{\dn}{^{\vphantom{\dagger}}}

\newcommand{\ra}{\rightarrow}
\newcommand{\la}{\leftarrow}
\newcommand{\ua}{\uparrow}
\newcommand{\da}{\downarrow}
\newcommand{\bb}[1]{\mathbb{#1}}

\newcommand{\eps}{\epsilon}

\newcommand{\tpsi}{\tilde{\psi}}

\newcommand{\sgn}[1]{{\rm sign}{#1}}
\newcommand{\pref}[1]{(\ref{#1})}

\newcommand{\intoinf}[1]{\int_{0}^{\infty}{#1}}

\newcommand{\abs}[1]{\left\vert #1 \right\vert}

\newcommand{\braket}[1]{\left\langle #1\right\rangle}

\newcommand{\com}[2]{\left[#1,#2\right]}
\newcommand{\acom}[2]{\left\{#1,#2\right\}}
\newcommand{\mat}[1]{\left(\bmx{cc}#1\emx\right)}
\newcommand{\matc}[2]{\left(\bmx{#1}#2\emx\right)}

\newcommand{\bw}[1]{\begin{widetext}}
\newcommand{\ew}[1]{\end{widetext}}

\newcommand{\red}[1]{{#1}}
\newcommand{\gray}[1]{}%{{\small \color{gray} #1}}

\begin{document}
\title{\red{Isolating Kondo Anyons} for Topological Quantum Computation}
\author{Yashar Komijani$^{1, *}$}
 \affiliation{ $^1$Department of Physics and Astronomy, Rutgers University, Piscataway, New Jersey, 08854, USA}
\date{\today}
\begin{abstract}
We propose to use residual anyons of the overscreened Kondo effect for topological quantum computation. A superconducting proximity gap of $\Delta< T_K$ can be utilized to isolate the anyon from the continuum of excitations and stabilize the non-trivial fixed point. We use weak-coupling renormalization group, dynamical large-$N$ technique and bosonization to show that the residual entropy of multichannel Kondo impurities survives in a superconductor. We find that while (in agreement with recent numerical studies) the non-trivial fixed point is unstable against intra-channel pairing, it is robust in presence of a finite inter-channel pairing. Based on this observation, we suggest a superconducting charge Kondo setup for isolating and detecting the Majorana fermion in the two-channel Kondo system.
\end{abstract}
\maketitle
\emph{Introduction} - 
The quest for realizing non-Abelian anyons, like Majorana zero modes (MZMs) and para-fermions, has led to an extensive research due to their application in topological quantum computation \cite{NayakRMP,Alicea2019}. However, in spite of a decade of active research and considerable progress, an unequivocal demonstration of MZMs in non-interacting systems is yet to be seen. Moreover, the currently pursued Ising anyons are insufficient for an all-topological quantum computation, which requires Fibonacci anyons. The main option for realizing parafermions is the edge state\red{s} of fractional quantum Hall \red{systems} in proximity to a superconductor (SC)\,\cite{Mong14,Vaezi14}. Even so, an elaborate technique is required to isolate the Fibonacci sector\,\cite{Hu18}.  Here, we propose an alternative route of using the fractionalization inherent to the Kondo effect, to realize MZMs and Fibonacci anyons.  

The Kondo effect arises when the electrons in a metal screen a magnetic impurity spin (Fig.\,1a), so that the spin effectively disappears at low temperatures\,\cite{AndreiRMP}. When various channels compete in screening a magnetic impurity, the spin is overscreened and this typically leads to a fractionalization of the spin and a residual degree of freedom at low temperatures\,\cite{Nozieres80,Andrei84,
EmeryKivelson,Affleck93,Coleman1995,Bulla08}. In the simplest case of two-channel Kondo (2CK) model, the infrared (IR) fixed point (FP) contains a decoupled MZM with ground state degeneracy of $g_{\rm 2CK}=\sqrt 2$, similar to the edge mode of the 1D Kitaev model. The three channel Kondo (3CK) has  $g_{\rm 3CK}=(1+\sqrt 5)/2$, corresponding to a Fibonacci anyon. Can these anyons, e.g. the 2CK MZM, be utilized for quantum computation? 

Nowadays, multichannel Kondo systems are not as out-of-reach as before. Experiments on semiconducting quantum dots\,\cite{Potok2007,Keller2015} and charge Kondo effect\,\cite{Matveev91,Furusaki95a,Furusaki95} in quantum Hall regime\,\cite{Iftikhar2015,Iftikhar2018} have demonstrated 2CK and 3CK physics and there are also other proposals based on Majorana boxes,\,\cite{Altland14} Floquet-driven Anderson impurity\,\cite{Floquet}, or magnetically frustrated Kondo systems\,\cite{Affleck95,Ingersent05,KoenigKomijani}. Even though a MZM can be achieved using simpler non-interacting setups, a 2CK realization paves the way for producing Fiboancci anyons using 3CK \red{physics}. However there are various obstacles:

i) It is unclear if the residual MZM in the 2CK model is localized at the position of the impurity spin, or delocalized throughout the system.

ii) Local relevant spin/channel symmetry-breaking perturbations destablize the non-trivial fixed point\,\cite{AffleckLudwigCox92}.

iii) The gapless spectrum of the conduction band prohibits singling-out the topological sector and braiding.

iv) The coupling between two spin-impurities mediated by the conduction band destabilizes individual 2CK FPs, driving them to a Fermi liquid at the IR\,\cite{Georges1997}.

\begin{figure}[tp!]
\includegraphics[width=\linewidth]{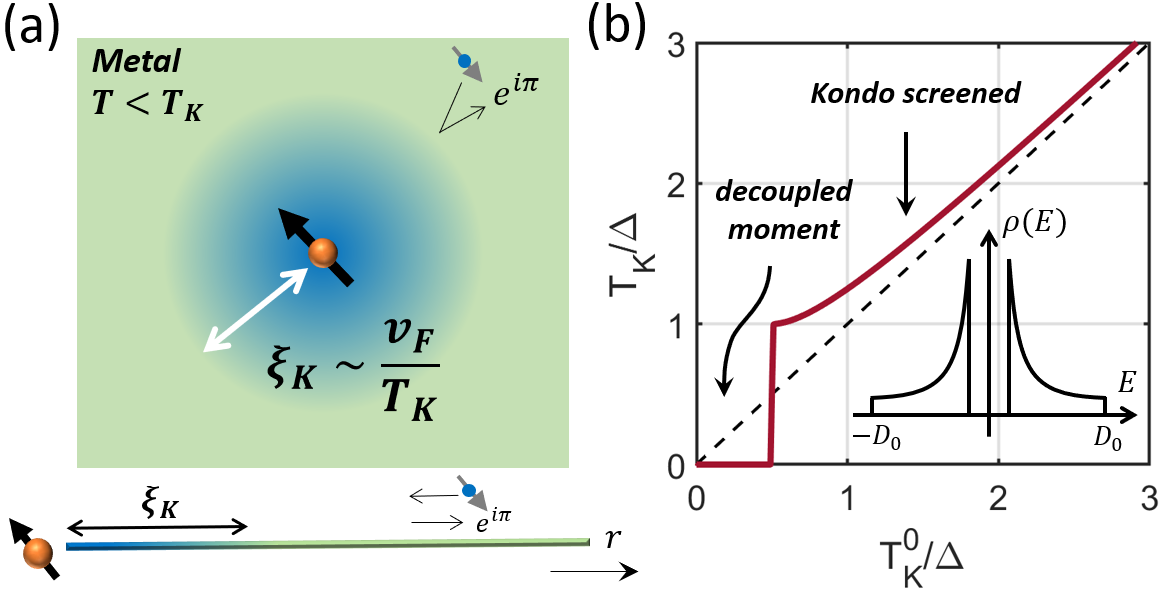}
\caption{\small\raggedright (a) A Kondo impurity in a metal is screened within a length scale of Kondo cloud\,\cite{Affleck2010} so that other electrons only experience a $e^{2i\delta_c}$ with $\delta_c=\pi/2$ phase shift. Bottom: The problem can be reduced to a 1D problem in the radial direction. (b) $T_K(\Delta)$ as a function of $T_K^0$. There is a threshold $T_K^0=\Delta/2$ below which, the system remains in the local moment $S=\log(2)$ phase. The inset shows the local density of states in the superconducting region.
 }
\end{figure}

To solve these problems, we suggest to gap out the conduction band at low-energies by a superconductor with a gap $\Delta<T_K$. The low-energy effective theory of such a system is only in terms of fractionalized degrees of freedom  $\gamma_j$ localized at the position of spin impurities, suitable for braiding. For example in the 2CK case,  $H_{\rm eff}=i\sum_{m<n}M_{mn}\gamma_m\gamma_n+{\cal O}(\gamma^4)$.
The spin impurities do not need to move in real space and a braiding in configuration space of $M_{mn}$ is sufficient for computation. 

For simplicity and practicality we consider spin-singlet s-wave pairing.
% A 2D problem can be mapped to a semi-infinite 1D wire with the Kondo impurity at the end \cite{SM}. 
For \red{an} infinitesimal Kondo coupling, the local moment remains decoupled and \red{it remains so}, as long as $T_K<\Delta$ \cite{Balatsky,Pasnoori19}. On the other hand, the $T_K>\Delta$ regime is expected to connect smoothly to the $\Delta\to 0$ limit. Indeed, for a single-channel Kondo impurity at the strong-coupling FP one can ``pair" Kramers pairs of $\pi/2$ phase-shifted IR quasi-particles.

However, at the non-Fermi liquid ground state of overscreend Kondo systems, an in-going single-particle state is scattered into out-going many-body states\,\cite{Affleck93} (no single-particle amplitude in the 2CK case). In absence of well-defined quasi-particles, it is unclear whether it is possible at all to open up a gap and how the non-trivial FP can smoothly connect to a superconducting state far away from the impurity. 
In fact recent numerical renormalization group (NRG) studies have indicated that the 2CK FP is unstable against pairing\,\cite{Zitko17}.

Considering the potential application, we revisit the problem here. After a brief discussion of weak-coupling regime, we use dynamical large-$N$ technique to show that the residual entropy survives in the limit of a finite $\Delta<T_K$. To gain insight about the more relevant SU(2) spin, we  use bosonization. We find that although the 2CK FP is unstable against intra-channel pairing, it is robust against inter-channel pairing, and we propose a setup to isolate the MZM. 

\emph{Model} -
The model consists of $K$ channels of non-interacting spinful electrons, proximity\red{-}paired to a SC and Kondo coupled to an impurity spin. The Hamiltonian is $H=H_0+H_\Delta+H_K$, where
\be
H_0=\sum_{k,a,\alpha}\eps_kc\dg_{ka\alpha}c\dn_{k\alpha}, \quad H_K=J_K\sum_{\red{kk',}a,\alpha\beta}\vec S\cdot c\dg_{ka\alpha}\vec\sigma_{\alpha\beta}c\dn_{k'a\beta}\label{eq1}
\ee
Here, $a=1\dots K$ is the channel index. \red{In the SU(2) case $\vec S$ is a $S=1/2$ spin operator, $\vec\sigma$ are the Pauli matrices and $\alpha,\beta=\ua,\da$}. We consider a singlet proximity pairing
\be
H_{\Delta}=\Delta\sum_{\red{k,}a} [c\dg_{k,a,\ua}c\dg_{-k,a,\da}+h.c.],\label{eq2}
\ee
which for a wide conduction band of electrons, $\abs{\eps_k}<D_0$, results in the local  Green's function (without Kondo)
\be
g_c(z)=-2\rho \frac{z\tau^0+\Delta\tau^x}{\sqrt{z^2-\Delta^2}}\log\frac{D_0-\sqrt{z^2-\Delta^2}}{-D_0-\sqrt{z^2-\Delta^2}}
\ee
with $\tau^{0/x}$ being Pauli matrices in the Nambu space. 

\emph{Weak-coupling} - At the weak-coupling \red{limit} (small $J_K/D_0$) for any $K$, the Kondo coupling evolves as $dJ_K/d\ell=\rho(D)J_K^2$ where $d\ell=-dD/D$ and $\rho(\eps)=g''_{c,ee}(\eps-i0^+)$. This density of states is shown in the inset of Fig.\,1b. As the cut-off $D$ is reduced $J_K(D)$ increases. $T_K(\Delta)$ is defined as the $D$ at which $J_K$ diverges. If this happens before $D$ is reduced to $\Delta$, the Kondo coupling has already renormalized to its infrared value and the moment is fully/over-screened. Otherwise, the ground state is an unscreened local moment, separated from the screened phase with a first-order transition\,\cite{Balatsky}. The $T_K(\Delta)$ as \red{a} function of $T_K^0=D_0e^{-2D_0/J_K}$ is shown in Fig.\,1b. The two phases are separated by $T_K^0/\Delta= 1/2$. In order to explicitly study the infrared fixed point, we use the large-$N$ limit.

\emph{Dynamical large-$N$} - We use Schwinger bosons\,\cite{ParcolletGeorges97} to represent SU(N) spins as $S_{\alpha\beta}=b\dg_\alpha b\dn_\beta$ with $\alpha,\beta=1\dots N$. The bosons have larger Hilbert space than the original spin and a constraint $b\dg_\alpha b\dn_\alpha=2S$ has to be imposed to stay within the physical subspace. Plugging this into $H_K$ and using a Hubbard-Stratonovitch transformation, $H_K$ reduces to
\be
H_K=\sum_a\frac{\chi\dg_a\chi_a}{J_K}+\sum_{a\alpha}[\chi\dg_ab\dn_\alpha c\dg_{a\alpha}+h.c.],
\ee
where Grassmannian ``holons'' $\chi$ are introduced. Keeping the ratios finite, but taking the $N,K,2S\to\infty$ limit, the Green's functions and self-energies of bosons and holons obey simple forms:
\be
\Sigma_B(\tau)=-\frac{K}{N}g_{c,ee}(\tau)G_\chi(\tau), \quad \Sigma_\chi(\tau)=g_{c,ee}(-\tau)G_B(\tau),
\ee
which can be solved self-consistently together with the Dyson equations $G_B^{-1}(z)=z-\lambda-\Sigma_B(z)$ and $G_\chi^{-1}(z)=-J_K^{-1}-\Sigma_\chi(z)$. See Refs.\,\cite{ParcolletGeorges97,Coleman2005,Rech2006,Komijani18a} for details.
\begin{figure}[tp!]
\includegraphics[width=\linewidth]{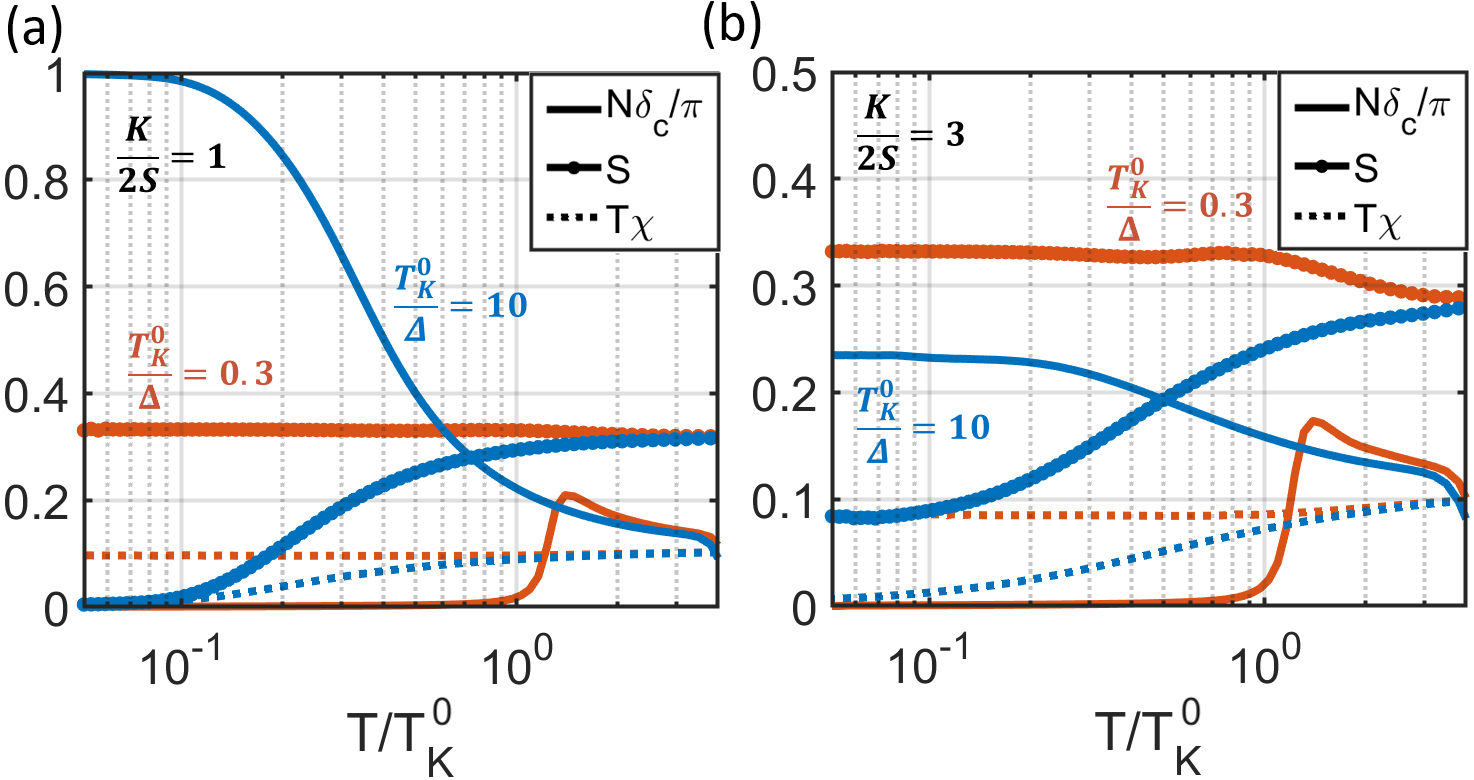}
\caption{\small\raggedright Conduction electron phase shift $N\delta_c/\pi$, thermodynamical entropy $S$ and effective moment $T\chi$ as function of $T/T_K^0$ for (a) fully screened case $K/2S=1$ and (b) overscreened case $K/2S=3$. Blue curves correspond to $T_K^0/\Delta=10$ and red curves correspond to $T_K^0/\Delta=0.3$. The calculation has been done for $2S/N=0.1$.
 }
\end{figure}

We first study the fully screened case $K=2S$. The  phase shift $N\delta_c$, the residual entropy $S$, and the effective moment $T\chi$ are shown in Fig.\,2a as a function of $T/T_K^0$. The local moment phase at $T_K^0/\Delta<1$ (red) and the screened phase at $T_K^0/\Delta>1$ (blue) are clearly visible. Fig.\,2b shows the same quantities in the overscreened case. While for $T_K^0/\Delta>1$ (blue), the moment disappears at low temperature, the residual entropy (\red{same value as} the gapless system\,\cite{Andrei84,Affleck93}) survives. \red{However, it remains unclear whether this large-$N$ result} holds for $N=K=2$ and $2S=1$. Therefore, in order to better examine the SU(2) case, we use field-theory techniques.

\emph{Bosonizaton} - The Hamiltonian $H$ in 2D can be reduced to a sum of 1D wires (in radial direction) terminated at the position of the impurity (Fig.\,3, Apppendix \ref{refA}). We first linearize the spectrum using $c_{a\alpha}=e^{ik_Fx}\psi_{Ra\alpha}+e^{-ik_Fx}\psi_{La\alpha}$ and use the recipe $\psi_{L/R,a\alpha}=F_{a\red{\alpha L/R}}\exp[{i\sqrt\pi(\phi_{a\alpha}\pm\theta_{a\alpha})}]$ to express the left/right-moving fermions in terms of conjugate bosons with commutation relations $[\phi_{a\alpha}(x),\theta_{b\beta}(y)]=-\frac{i}{2} \sgn(x-y)\delta_{ab}\delta_{\alpha\beta}$ \red{and mutually anti-commuting charge-lowering operators $F_{a\alpha}$, referred to as Klein factors\,\cite{Zarand00}}. The UV open boundary condition at the position of the impurity corresponds to $\theta_{a\alpha}(0)=0$ \red{and $F_{a\alpha R}=F_{a\alpha L}=F_{a\alpha}$.}

\begin{figure}[tp!]
\includegraphics[width=\linewidth]{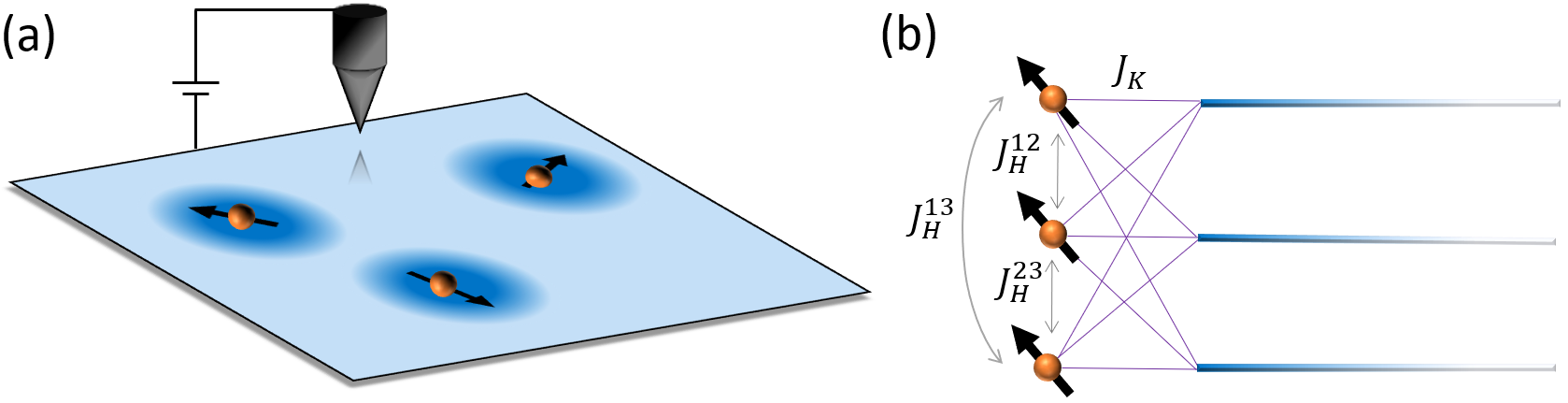}
\caption{\small\raggedright (a) The problem of three (multichannel) Kondo impurities in 2D can be mapped to (b) the problem of three Kondo impurities at the end of semi-infinite wires. This is so even in presence of proximity pairing of the host metal. Also various potential scatterings (e.g. a scanning gate microscope) only modify the mutual Kondo coupling and RKKY interactions among the impurities.}
\end{figure}

\emph{1CK} -  In the single-channel case ($K=1$) we define charge/spin bosons according to $\phi_{c/s}=(\phi_{\ua}\pm\phi_{\da})/\sqrt 2$ (and same for $\theta_{c/s}$) in terms of which
\be
H_0=\frac{v_F}{2}\sum_{\mu=c,s}\intoinf{dx[(\partial_x\phi_\mu)^2+(\partial_x\theta_\mu)^2]},
\ee
and the proximity pairing becomes \red{(${\cal I}^+\equiv F\dg_\ua F\dg_\da$)}
\be
\hspace{-.114cm}H_{\Delta}\sim\frac{\Delta}{\pi}\intoinf{dx}\cos\sqrt{2\pi}\theta_s(x)[\red{{\cal I}^+} e^{-i\sqrt{2\pi}\phi_c}+h.c.].\label{eqHDel}
\ee
\red{The acompanying $e^{-i\sqrt{2\pi}\phi_c}$ factor ensures that ${\cal I}^+$ always appears with ${\cal I}^-$ in the partition function. Whenever this is the case, bosonic combination of Klein factors, e.g. ${\cal I}^+$ can be replaced by 1 and fermionic ones by a corresponding Majorana fermion.}
Thus, Eq.\,\pref{eqHDel} leads to the pinning $\phi_c=0,\theta_s=\sqrt{\pi/2}$ or vice versa at the \red{IR}. An anisotropic Kondo interaction bosonizes to
\be
\hspace{-.14cm}H_K=\frac{J_K^\perp}{2}[S^+\red{{\cal S}^-} e^{i\sqrt{2\pi}\phi_s(0)}+h.c.]-\frac{J_K^z}{\sqrt{2\pi}}S^z\partial_x\theta_s(0).
\ee
\red{where ${\cal S}^-\equiv F\dg_\da F\dn_\ua$.}
We use the unitary transformation $U=\exp[i\mu S^z\phi_s(0)]$ in order to tune the system $H\to U\dg HU$ to the Toulouse II\,\cite{KotliarSi} or \emph{decoupling}\,\cite{Komijani15} point. 
%This is best achieved by unfolding the system in terms of  chiral right-movers $\phi,\theta(x)=[\varphi(-x)\pm\varphi(x)]/\sqrt{2}$. 
For any $K$, the decoupling point is defined as the choice of $\mu$ which maximizes the scaling dimension for the transverse Kondo coupling $J_K^\perp$\,\cite{KomijaniKoenig}. For the single channel case, the bosonic factor $e^{i\sqrt{2\pi}\phi_s}$ can be eliminated by $\mu=\sqrt{2\pi}$. Moreover, $\tilde\phi(x)=\phi(x)$ but
\be
\tilde\theta(x)=\theta(x)+\mu\sqrt{2} S^z\sgn(x),\label{eq9}
\ee
and the Hamiltonian becomes 
\be
U\dg (H_{K}+H_0)U=H_0+ \frac{J_K^\perp}{2} [S^+\red{{\cal S}^-}+h.c.]-\frac{\tilde J^z_K}{\sqrt 2\pi}\partial_x\theta_s(0)S^z.\nonumber
\ee
At the decoupling point, $J_K^\perp$ is highly relevant whereas the term proportional to $\tilde J^z_K=J_K^z-2\mu\sqrt \pi v_F$ is (marginally) irrelevant. Via this term the conduction electrons `observe' the state of the spin \red{(along $z$-direction)} and entangle-to it, resulting in its decoherence. The strong coupling FP corresponds to $J_K^\perp\to\infty$ and $\tilde J_K^z\to 0$, \red{i.e. a singlet between the dressed spin $S$ and the Klein-factor composite ${\cal S}$.}

Now, we turn to coexisting Kondo and pairing terms.  Due to Eq.\,\pref{eq9} the pairing term is modified by the unitary transformation $U\dg H_\Delta U$, and the pinning of $\theta_s(x)$ depends on the state of \red{the spin}. An infinitesimal $J_K^\perp \vec S_\perp\cdot\vec{\cal S}_\perp$ cannot flip the spin. However, a large $J_K^\perp$ can `melt' the bosonic solid of $\theta_s(x)$ in an area $\xi_K$. Near the impurity, $\tilde J_K^z$ is large and the spin precession is negligible. But at long distances $\tilde J_K^z\to 0$ and the spin-precession is set by $J_K^\perp$ which grows at low energies, i.e. $S^z(\tau)=e^{\tau J^\perp\vec S_\perp\cdot\vec{\cal S}_\perp}S^z e^{-\tau J^\perp \vec S_\perp\cdot\vec{\cal S}_\perp}$ is fluctuating in time, with a rate given by $J^\perp$. For $\theta_s(x)$ to follow this evolution, it costs an energy $\frac{v_F}{2}[\partial_x\phi_s]^2=\frac{v_F}{2}[\partial_\tau\theta_s]^2\propto[\partial_\tau S^z(\tau)]^2$ per unit length. Hence beyond certain distance, $\theta_s(x)$ can no longer follow the spin and it is pinned to $\braket{S^z}\red{=0}$. This defines a characteristic distance $\xi_K\sim v_F/T_K(\Delta)$\,\cite{Affleck2010}.

\emph{2CK} - In the two-channel case $(K=2)$, we follow\,\cite{EmeryKivelson} and define collective bosons according to
\be
\mat{\phi_c \\ \phi_s \\ \phi_f \\ \phi_{sf}}=\frac{1}{2}\matc{rrrr}
{1 & 1 & 1 & 1 \\
1 & -1 & 1 & -1 \\ 
1 & 1 & -1 & -1 \\ 
1 & -1 & -1 & 1}
\mat{\phi_{1\ua} \\ \phi_{1\da} \\\phi_{2\ua} \\ \phi_{2\da}},\label{eq10}
\ee
and similarly for $\theta_\mu$. Again $U=\exp[i\mu S^z\phi_s(0)]$ with $\mu=\sqrt\pi$ transforms $H_K$ to
\be
\red{J_K^\perp [S^+{\cal F}\dn_s({\cal F}\dn_{sf}e^{i\sqrt\pi\phi_{sf}(0)}+h.c.)+h.c.]-\frac{\tilde J_K^z}{\sqrt \pi}S^z\partial_x\theta_s(0).}\nonumber
\ee
\red{We have defined four new Klein factors (Appendix \ref{refB}) in accord with Eq.\,\pref{eq10} and identified ${\cal S}^-_1={\cal F}\dn_s{\cal F}\dn_{sf}$  and ${\cal S}^-_2={\cal F}\dn_s{\cal F}_{sf}\dg$. Note that since $S^+$ always accompanies ${\cal F}_s$, the total spin fluctuates only mildly. Representing $S^+{\cal F}_s=d\dg$ with a Dirac fermion $d\equiv\gamma+i\gamma'$, and re-fermionizing the dim-1/2 operator $\psi_{L/R,sf}\equiv {\cal F}_{sf}e^{i\sqrt\pi[\phi_{sf}(x)\pm\theta_{sf}(x)]}$,} the Hamiltonian becomes a resonant Andreev scattering
\be
U\dg H_KU \to 2iJ_K^\perp\gamma'\eta_{sf}(0)-\frac{\tilde J_K^z}{\sqrt\pi}i\gamma\gamma'\partial_x\theta_s(0).\label{eq14}
\ee
\red{where $\eta_{sf}(x)\equiv \frac{1}{2}(\psi\dn_{sf}+\psi\dg_{sf})$. It is safe to represent ${\cal F}_{sf}$ with a Majorana fermion $\to\Gamma_{sf}$, here.} The ground state of $J_K^\perp$ term is thus a Schr\"odinger's cat state\,\cite{Affleck2013} between the spin and the pinning of the boson $\phi_{sf}(0)=0,\sqrt\pi$. This means the difference of the spin in the two channels, $\theta_{sf}(0)$, which is the only non-conserved charge, strongly fluctuates\,\cite{vonDelft98}. 
The first term of \pref{eq14} hybridizes $\gamma'$  with the conduction \red{Majoranas} $\eta_{sf}(0)$, and provides it with the scaling dimension 1/2. Near this IR FP, the originally marginal interaction $\tilde J_K^z$, coupling $\gamma$ and $\gamma'$, becomes a dim-3/2 irrelevant operator (the leading\,\cite{Sengupta94}). At the IR FP, $\gamma$ is entirely decoupled\,\cite{EmeryKivelson} consistent with the $S=\frac{1}{2}\log 2$ residual entropy,\,\cite{Andrei84,Affleck93} realizing a MZM. 

Can this MZM survive in a SC? The simplest form of SC is an induced intra-channel pairing (c.f. Eq.\,\ref{eqHDel}),
\bea
{ H}^{\rm intra}_{\Delta}=\frac{4\Delta}{\pi}\intoinf{dx}&\Big\{&\cos\sqrt{2\pi}\theta_{s1}(x)\cos\sqrt{2\pi}\phi_{c1}
\qquad\nonumber\\&+&
\cos\sqrt{2\pi}\theta_{s2}(x)\cos\sqrt{2\pi}\phi_{c2}\Big\},\qquad
\eea
\red{where we have set ${\cal I}^\pm_j\to 1$, as before}. $U\dg H_\Delta^{\rm intra}U$ has the effect of $\theta_s\to \tilde\theta_{s}$ as in Eq.\,\pref{eq9}. Far from the impurity the two lines can be minimized independently. Since both $\theta_{s1}$ and $\theta_{s2}$ are pinned, $\theta_{sf}=(\theta_{s1}-\theta_{s2})/\sqrt 2$ is also pinned. This means that $\phi_{sf}$ is strongly fluctuating and the term \red{$J_K^\perp i\gamma'\Gamma_{sf}\cos\sqrt\pi\phi_{sf}(0)$} becomes highly irrelevant. Thus inclusion of a small $\Delta$ destabilizes the 2CK FP, in agreement with NRG results.\,\cite{Zitko17}\\
 On the other hand, a singlet/triplet inter-channel pairing has the form
\red{
\bea
H_\Delta^{\rm inter}&=&\Delta\intoinf{dx[(c\dg_{1\ua}c\dg_{2\da}\pm c\dg_{2\ua}c\dg_{1\da})+h.c.}]\\
&\sim&\frac{\Delta}{\pi}\Big\{{\cal F}\dg_c e^{-i\sqrt\pi\phi_c}[{\cal F}\dg_{sf}e^{-i\sqrt\pi\phi_f}\cos\sqrt\pi(\theta_s+\theta_{f})\nonumber\\
&&\qquad\qquad\pm {\cal F}\dn_{sf}e^{i\sqrt\pi\phi_f}\cos\sqrt\pi(\theta_f-\theta_s)]+h.c.\Big\}\nonumber
\eea
using $F\dg_{1\ua}F\dg_{2\da}={\cal F}\dg_c{\cal F}\dg_{sf}$ and $F\dg_{1\da}F\dg_{2\ua}={\cal F}\dg_c{\cal F}\dn_{sf}$. Again, we can safely replace Klein factors with Majorna fermions  ${\cal F}_\mu\to\Gamma_\mu$ because of the exponential factors and} $U\dg H_\Delta^{\rm inter}U$ has the effect of $\theta_s\to \tilde\theta_{s}$ as in Eq.\,\pref{eq9}. Note that there is no $\theta_{sf}$ here! This interaction tends to pin $\phi_{sf}$ value and is \red{benign to the 2CK FP as the bulk pinning can} smoothly connect to the boundary value. \red{The harmlessness of the inter-channel pairing can also} be seen in a two-site problem, where the addition/removal of a pair of inter-channel electrons maps the strong coupling to weak-coupling or vice versa without affecting the channel isotropy. The possibility of coexistence of interchannel pairing and 2CK FP seen here, can be verified in future NRG studies.

\emph{Multiple impurities} - The case of two 2CK impurities coupled to the same bath was discussed before\,\cite{Georges1997}. The double 2CK FP is transformed to a line of FPs by the RKKY interaction $J_H\vec S_1\cdot\vec S_2$ which becomes a marginal operator $iJ'_H\gamma_1\gamma_2\partial_x\Phi(0)$ at the non-trivial FP. Here, $J'_H$ is the renormalized coupling and $\Phi$ is a linear combination of the spin bosons of each impurity\,\cite{Georges1997}; the two decoupled Majoranas form a non-local charge qubit whose state is dynamically measured (and decohered) by the gapless $\partial_x\Phi$ mode. Presence of a gap in the spectrum has the additional feature of suppressing such decoherence effects and reducing it to a static $iM_{12}\gamma_1\gamma_2$ interaction discussed before.

\emph{Experimental realization} - Based on above discussion, we propose a modified version of the charge Kondo setup\,\cite{Iftikhar2015,Iftikhar2018} at zero magnetic field to isolate the MZM in the 2CK case. In the simplest charge Kondo effect, a spinless single electron transistor (SET) with large charging energy is coupled to a spinless conduction bath. The SET is tuned to a charge degeneracy point $\Delta Q=0,1e$ so that the \emph{charge parity} plays the role of the pseudo-spin. The location of the electron, either in SET or the conduction bath, plays the role of the conduction electron pseudo-spin\,\cite{Matveev91}. The spinful case provides the simplest realization of a two-channel charge Kondo effect. Incidentally, this is ideally suitable for combining with previous discussion. A proximity pairing of the SET (or SET made of SC) and conduction bath with singlet SC leads to purely inter-channel pairing.\,\footnote{Since the role of the spin and channel are reversed, a singlet pairing in the lead and the SET correspond to $T_\pm$ inter-channel spin-triplet pairing in the original basis, whereas Eq. (10) showed that a $T_0$ triplet pairing is benign to 2CK physics. Considering $c\dg_{1\ua}c\dg_{2\ua}-c\dg_{1\da}c\dg_{2\da}=c\dg_{1\ra}c\dg_{2\la}-c\dg_{1\la}c\dg_{2\ra}$, we expect the latter to hold for all inter-channel triplet pairings albeit with a $\pi$ phase difference, although this cannot be seen in abelian bosonization.} Such Coulomb blockaded superconducting islands are common in topological Kondo effect\,\cite{Beri12,Altland14}. However, the MZM here is produced by the 2CK, rather than by the band topology.

Realization of the charge Kondo effect requires $\delta E\ll k_BT\ll T_K\ll E_C$ where $\delta E$ is the mean-level spacing, $T_K\sim E_Ce^{-\pi^2/2{\cal T}}$ is the Kondo temperature ex\red{p}ressed in terms of transmission ${\cal T}$, and $E_C=e^2/C$ is the charging energy. This condition can be met in small metallic grains, e.g. the hybrid metal-semiconductor setup of Iftikhar et al.\,\cite{Iftikhar2015,Iftikhar2018}. Alternatively, a purely proximity-induced superconductivity in semiconductor heterostructures with large carrier mass can be used. Since carrier mobility is unimportant, one possible option is a dot with a large charging energy defined using in-plane gates in a shallow 2D valence band hole gas.\, \cite{Komijani2008,Komijani13}

\emph{Detection} - The presence of the MZM has to be inferred indirectly; we consider an additional \emph{normal} lead (e.g. a scanning tunnelling microscope) \emph{weakly coupled} to the SET to measure the conductance between the two leads. At the 2CK FP, the coupling of the probe channel is irrelevant\,\cite{Furusaki95a,Furusaki95} and a conductance of $G\propto T$ is expected on resonance at $T>\Delta$.\,\cite{Furusaki95a,Furusaki95} For $T<\Delta$, the capacitive coupling to the SET, with another scanning SET \red{might be a possibility, but its feasibility requires further studies}. Alternatively, entropy measurements along\,\cite{Sela19} can be envisioned. 

\emph{Conclusion} - We have proposed to use Kondo-based anyons in proximity with superconductivity for quantum computation. We found that the residual entropy survives a gap in the spectrum, particularly if the gap is produced by an inter-channel proximity pairing. \red{The presence of the gap in the spectrum protects the Majorana fermion and the ground state degeneracy against small symmetry-breaking perturbations.} We have suggested a superconducting version of the charge Kondo setup for isolating the MZM in the 2CK model.

\emph{Acknowledgement} - Support from NSF grant DMR-1830707 is gratefully acknowledged. The author is indepted to P.~Coleman for insightful comments and the encouragement to write up this manuscript and to  E.~K\"onig for renewing his interest in the topic and collaborations.\, \cite{KoenigKomijani,KomijaniKoenig} It is a pleasure to acknowledge illuminating discussions with I.~Affleck, N.~Andrei, A.~Sengupta and Y.~Meir and valuable inputs from M.~Gershenson and K.~Matveev regarding the proposed experiment.

After posting this paper, we became aware of a recent paper\,\cite{Lopes19} that was appeared online few days before and also seeks to use multichannel Kondo anyons for computation, but in gapless sytems. The two papers however, do not have any significant overlap.
%--------------------------------------------------------------------------------------
%\emph{{Introduction} -} 
%--------------------------------------------------------------------------------------
\section*{Appendices}
In these appendices we provide supporting documents for some of the statements in the paper. Appendix \ref{refA} shows that the problem of $N$ multi-channel Kondo impurities immersed in 2D and 3D can be reduced to a set of one-dimensional problems, including the superconducting proximity pairing term. Appendix \ref{refB} summarizes our convention for the Klein factors that are used in bosonization.
\subsection{Dimensional reduction}\label{refA}
For completeness, here we show how a problem of $M$ impurities in two dimension can be reduced to a sum of 1D metallic systems terminated at the spin impurities, appropriate for field theory analysis. This is a generalization of $M=2$ case from Ref.\,\cite{Affleck95}. We start from a two-dimensional system described by $H=H_0+H_K+H_\Delta$ defined in Eqs.\,\ref{eq1} and \ref{eq2}, and generalize it to the case that there are $M$ Kondo impurities located at positions $\vec d_n$ coupled to the conduction band.

\subsubsection{Kondo coupling}
We consider $M$ impurities located on a 2D plane at positions $\vec d_n$ shown in Fig.\,3. We have
\be
H_K=\int{\frac{d^2kd^2k'}{(2\pi)^4}}\psi\dg_{\vec k}\frac{\vec\sigma}{2}\psi\dn_{k'}\cdot\sum_{n=1}^MJ_n\vec S_n e^{-i(\vec k-\vec k').\vec d_n}.
\ee
We start by defining the fermions
\be
\psi_{nE}=\int{\frac{d^2k}{(2\pi)^2}}\delta(\eps_k-E)e^{i\vec k\cdot\vec d_n}\psi\dn_{\vec k},
\ee
in terms of which the Kondo Hamiltonian is
\be
H_K=\int{dEdE'}\sum_{n=1}^MJ_n\psi\dg_{nE}\frac{\vec\sigma}{2}\psi\dn_{nE'}\cdot\vec S_n.
\ee
The problem is that $\psi_{nE}$ do not obey standard anticommutation relations. Rather
\be
\acom{\psi\dn_{nE}}{\psi\dg_{mE'}}=g_{nm}(E)\delta(E-E'),
\ee
where
\be
g_{nm}(E)=\int{\frac{d^2k}{(2\pi)^2}}\delta(\eps_k-E)e^{i\vec k\cdot(\vec d_n-\vec d_m)}.
\ee
Representing the vector $\vec d_{nm}\equiv \vec d_n-\vec d_m$ in polar coordinates by $\vec d_{nm}=d_{nm}(\cos\phi_{nm},\sin\phi_{nm})$ we can write
\bea
g_{nm}(E)&=&\frac{k_E}{2\pi\partial_k\eps_E}\int_0^{2\pi}\frac{d\phi}{2\pi}e^{ik_Ed_{nm}\cos(\phi-\phi_{nm})}\nonumber\\
&=&\frac{k_E}{2\pi\partial_k\eps_E}\sum_pe^{-ip\phi_{nm}}J_p(k_Ed_{nm})i^p\int_0^{2\pi}\frac{d\phi}{2\pi}e^{ip\phi}\nonumber\\
&=&g_EJ_0(k_Ed_{nm}),
\eea
where
\be
g_E=\frac{k_E}{2\pi\partial_k\eps_E}.
\ee
Remarkably, $g_{nm}$ depends only on mutual distances of the impurities, measured by the corresponding wavelength $k_Ed_{nm}$. The matrix $g_{nm}$ is real and symmetric and it has 
real eigenvalues $\lambda_n$ and can be diagonalized by orthogonal eigenvectors $\vec u_n$ where $g\vec u_n=\lambda_n\vec u_n$. So we can write $g_{nm}=\sum_p u_{np}\lambda_{p}u^*_{mp}$ and the orthogonality is $\sum_pu_{np}u^*_{mp}=\delta_{nm}$. Defining new operators with
\be
\psi_{nE}=\sum_i u_{ni}(E)\sqrt{\lambda_i(E)}\tpsi_{iE},
\ee
and
\be
\tpsi_{iE}\equiv\frac{1}{\sqrt{\lambda_i(E)}}\sum_n{u^*_{ni}}(E)\psi\dn_{nE}
\ee
we find that they are orthonormal
\be \acom{\tpsi_{iE}}{\tpsi_{jE'}}=\delta_{ij}\delta({E-E'}),
\ee
and the Kondo Hamiltonian becomes
\be
H_K=\int{dEdE'}\sum_{nij}J_{nij}(E,E')\vec S_n\cdot \tpsi\dg_{iE}\frac{\vec\sigma}{2}\tpsi\dn_{jE'}, 
\ee
where
\be
J_{nij}(E,E')=J_n\sqrt{\lambda_i(E)\lambda_j(E')}u_{ni}(E)u^*_{nj}(E').
\ee
Although the $J_{nij}$ couplings are complex in general, the relation $J^*_{nij}(E,E')=J_{nji}(E',E)$ ensures the hermiticity of the Hamiltonian. Whether or not a spin $n$ couples the channels $i$ and $j$ depend on the product of wavefunctions of $i$ and $j$ at the site $n$, which can be tuned by moving potential scatterings induced by scanning tips (Fig.\,3). Next, we do a Taylor expansion of $J_{nij}(E,E')$ function around Fermi energy and keep only the leading relevant term. 
$H_K$ can be written in the matrix form $[\bb J_n]_{ij}=J_{nij}$,
\be
H_K=\sum_n  \Psi\dg(0)[\bb J_n\vec S_n\cdot\frac{\vec \sigma}{2}]\Psi\dn(0).
\ee
where $\Psi$ is a vector in $m=1...M$ index and the spin is implicit. As we see the impurities talk to all the channels and scatter electron between all the channels. A result of the Taylor expansoin of $J_{nij}(E,E')$ is that RKKY interactions will be induced between the spins
\be
H_K\to H_K+\sum_{ij}J_H^{ij}\vec S_i\cdot\vec S_j.
\ee
\subsubsection{Mapping to 1D}
The mapping to (unfolded) left-movers follows Ref.\,\cite{Affleck95}
\be
\psi_L(x,t)=\frac{1}{\sqrt v}\int_{-D}^D dE e^{-iE(t+x/v)}\psi_E,
\ee
for $x\in(-\infty,+\infty)$ 
with commutation relations
\be
\{\psi\dg_L(x),\psi\dn_L(y)\}=2\pi\delta(x-y).
\ee
Alternatively, we can work in the folded space of left and right-movers defined as
\be
\psi_R(x,t)=\psi_L(-x,t), \qquad (x>0).
\ee
\subsubsection{Kinetic term}
As we saw above, the Kondo interaction involves $M$ 1D conduction bands pulled out of the 2D conduction band. The natural guess for their kinetic energy  is
\bea
H_0(\vec d)&\equiv&\int{dE E\psi\dg_{nE}\psi\dn_{nE}}\\
%&&\hspace{-1.5cm}=\int{dE}E\int\frac{d^2kd^2k'}{(2\pi)^4}\delta(\eps_k-E)\delta(\eps_k'-E)e^{i(\vec k'-\vec k)\cdot \vec d}\psi\dg_{\vec k}\psi\dn_{\vec k'}\nonumber\\
&=&\int\frac{d^2kd^2k'}{(2\pi)^4}\eps_k\delta(\eps_k-\eps_k')e^{i(\vec k'-\vec k)\cdot\vec  d}\psi\dg_{\vec k}\psi\dn_{\vec k'}.\nonumber
\eea
%I don't like the $\delta(\eps_k-\eps_k')$ part. 
Summing over all the positions we have
\bea
\int{d^2r}H_0(\vec r)
%&=&\int\frac{d^2kd^2k'}{(2\pi)^4}\eps_k\delta(\eps_k-\eps_k')\psi\dg_{\vec k}\psi\dn_{\vec k'}\int{d^2r}e^{i(\vec k'-\vec k)\cdot\vec  r}\nonumber\\
&=&\delta(0)\int\frac{d^2kd^2k'}{(2\pi)^2}\eps_k\psi\dg_{\vec k}\psi\dn_{\vec k},
\eea
which is the total Hamiltonian up to a $\delta(0)$, i.e. the total volume. Transformation from $\psi_{nE}$ to $\tpsi_{iE}$ is a unitary transformation at each energy and it doesn't change the kinetic part. Therefore, we have
\be
H_0\to\sum_{n=1}^M\int{dE E\tpsi\dg_{nE}\tpsi\dn_{nE}}
\ee
\subsubsection{Pairing term}
Without loss of generality, we consider a channel-diagonal s-wave singlet pairing
\be
H_\Delta=\int{\frac{d^2k}{(2\pi)^2}}(\Delta\psi\dg_{k\ua}\psi\dg_{-k\da}+h.c.).
\ee
Let us look at 
\bea
H_\Delta^+(\vec d_n)&\equiv&\Delta\int{dE}\psi\dg_{nE\ua}\psi\dg_{n,E\da}\nonumber\\
&&\hspace{-1.75cm}=\Delta\int{dE}\int{\frac{d^2kd^2k'}{(2\pi)^4}}\delta(\eps_k-E)\delta(\eps_{k'}-E)\psi\dg_{k\ua}\psi\dg_{q\da}e^{-i(\vec k+\vec k')\cdot \vec d_n}\nonumber
\eea
Summing over all the points gives a $(2\pi)^2\delta(\vec k+\vec k')$ term that  gives
\be
\int{d^2r}H^+_\Delta(\vec r)=\delta(0)\Delta\int{\frac{d^2k}{(2\pi)^2}}\psi\dg_{k\ua}\psi\dg_{-k\da},
\ee
which is our initial Hamiltonian up to the system volume. Therefore,
\bea
H_{\Delta}&\to &\Delta\sum_{n=1}^M\int{dE}(\psi\dg_{nE\ua}\psi\dg_{nE\da}+h.c.)\nonumber\\
&=&\Delta\sum_{ij}\int{dE}(\tpsi\dg_{iE\ua}\tpsi\dg_{jE\da}+h.c.).
\eea
\subsection{Klein factors}\label{refB}
Here, we briefly review the treatment of Klein-factors following Ref.\,\cite{Zarand00}. Denoting the total number of fermions $\mu$ with
\be
N_\mu=\intoinf{dx}c\dg_{\mu}(x)c\dn_{\mu}(x), \quad \mu\in\{1\ua,1\da,2\ua,2\da\},
\ee 
the Klein factors have the property (no summation):
\bea
&&F\dn_\mu F\dg_\mu=F\dg_\mu F\dn_\mu=1\\
&&\{F\dn_\mu,F\dg_{\nu}\}=2\delta_{\mu\nu}\\
&&\{F\dn_\mu,F\dn_\nu\}=0, \quad \forall \mu\neq \nu\\
&&\com{F_\mu}{N_\nu}=\delta_{\mu\nu}F_\nu\\
&&\com{F_\mu}{\phi_\nu(x)}=\com{F_\mu}{\theta_\nu(x)}=0.
\eea
In the paper we transform into collective boson basis:
\bea
\mat{{\cal N}_c \\ {\cal N}_s \\ {\cal N}_f \\ {\cal N}_{sf}}=\frac{1}{2}\matc{rrrr}
{1 & 1 & 1 & 1 \\
1 & -1 & 1 & -1 \\ 
1 & 1 & -1 & -1 \\ 
1 & -1 & -1 & 1}
\mat{N_{1\ua} \\ N_{1\da} \\N_{2\ua} \\ N_{2\da}}.
\eea
In accord with this transformation, we define new Klein factors ${\cal F}_\mu$ for $\mu=c,s,f,sf$ that satisfy 
\bea
&&{\cal F}\dn_\mu {\cal F}\dg_\mu={\cal F}\dg_\mu {\cal F}\dn_\mu=1\\
&&\{{\cal F}\dn_\mu,{\cal F}\dg_{\nu}\}=2\delta_{\mu\nu}\\
&&\{{\cal F}\dn_\mu,{\cal F}\dn_\nu\}=0, \quad \forall \mu\neq \nu\\
&&\com{{\cal F}_\mu}{{\cal N}_\nu}=\delta_{\mu\nu}{\cal F}_\nu\\
&&\com{{\cal F}_\mu}{\phi_\nu(x)}=\com{{\cal F}_\mu}{\theta_\nu(x)}=0.
\eea
It can be seen that these any off-diagonal relation combination, like $F\dg_\mu F\dn_\nu$ change the population of the collective modes ${\cal N}_\zeta$ by integers and thus can be represented by a combination of new Klein factors\,\cite{Zarand00}. For example,
\be
F\dg_{1\ua}F\dn_{1\da}={\cal F}\dg_{sf}{\cal F}\dg_s, \qquad F\dg_{2\ua}F\dn_{2\da}={\cal F}_{sf}{\cal F}\dg_s.
\ee 
In the paper, we repeatedly use these and similar relations.

\bibliography{Kondo}

\end{document}